\documentclass[12pt,preprint]{aastex}

\begin{document}

\title{On the Excess Dispersion in the Polarization Position Angle of Pulsar 
Radio Emission}
\author{Mark M. McKinnon}
\affil{National Radio Astronomy Observatory\altaffiltext{1}{The National
Radio Astronomy Observatory is a facility of the National Science
Foundation operated under cooperative agreement by Associated 
Universities, Inc.}, Socorro, NM \ \ 87801\ \ USA}

\begin{abstract}

  The polarization position angles (PA) of pulsar radio emission occupy
a distribution that can be much wider than what is expected from the 
average linear polarization and the off-pulse instrumental noise. Contrary 
to our limited understanding of the emission mechanism, the excess 
dispersion in PA implies that pulsar PAs vary in a random fashion. An 
eigenvalue analysis of the measured Stokes parameters is developed to 
determine the origin of the excess PA dispersion. The analysis is applied 
to sensitive, well-calibrated, polarization observations of PSR B1929+10 
and PSR B2020+28. The analysis clarifies the origin of polarization 
fluctuations in the emission and reveals that the excess PA dispersion 
is caused by the isotropic inflation of the data point cluster
formed from the measured Stokes parameters. The inflation of the cluster 
is not consistent with random fluctuations in PA, as might be expected 
from random changes in the orientation of the magnetic field lines in 
the emission region or from stochastic Faraday rotation in either the 
pulsar magnetosphere or the interstellar medium. The inflation of the 
cluster, and thus the excess PA dispersion, is attributed to randomly 
polarized radiation in the received pulsar signal. The analysis also 
indicates that orthogonal polarization modes (OPM) occur where the 
radio emission is heavily modulated. In fact, OPM may only occur where 
the modulation index exceeds some critical value, $\beta_c\simeq 0.3$.

\end{abstract}


\section{INTRODUCTION}

  The linear polarization of pulsar radio emission is generally attributed 
to charged particles streaming along dipolar magnetic field lines above the 
pulsar's polar cap (Radhakrishnan \& Cooke 1969). Given the pulsar's enormous 
magnetic field strength ($\approx 10^{12}$ gauss) and the stability of its 
magnetic field structure, as implied by the long term stability of pulsar 
average profiles (Helfand, Manchester, \& Taylor 1975), one might expect 
the polarization position angle (PA) at a particular pulse longitude to be 
fixed rigidly on the plane of the sky. However, PA histograms constructed from 
single-pulse polarization observations are much broader than what is expected
from the average linear polarization and the off-pulse instrumental noise 
(Manchester, Taylor, \& Huguenin 1975, hereafter MTH; Stinebring et al. 1984, 
hereafter SCRWB; McKinnon \& Stinebring 1998, hereafter MS1; Karastergiou et 
al. 2002), suggesting that pulsar PAs also vary in a random fashion.

  This excess dispersion in PA may be intrinsic to the pulsar, possibly 
arising from an analog of stochastic Faraday rotation in the pulsar 
magnetosphere or from randomly polarized radiation (RPR) in the emission. 
Since the orientation of a pulsar's linear polarization vector is thought to 
be aligned with the magnetic field lines in the emission region, the excess 
PA dispersion could be due to fluctuations in field line orientation. However, 
this explanation seems highly unlikely because it requires the orientation of 
a $10^{12}$ gauss magnetic field to fluctuate on a timescale comparable with 
a pulsar's rotation period ($P\simeq 1$s). Stochastic Faraday rotation in the 
interstellar medium (ISM) also seems an unlikely explanation for the origin of 
the excess PA dispersion because the rotation measure (RM) of the ISM does not 
fluctuate on the timescale of a rotation period. Furthermore, RM fluctuations 
should cause PA variations to be similar everywhere within a pulsar's pulse, 
but the PA dispersion appears to vary with pulse longitude. Whatever the cause 
of the excess dispersion, determining it requires a detailed understanding of 
what causes polarization fluctuations. To discriminate between mechanisms that 
create polarization fluctuations, analytical tools must be developed for a 
thorough interpretation of single-pulse polarization observations.

   The statistical model of radio polarimetry developed by McKinnon (2003b)
provides a foundation for the detailed interpretation of polarization
observations. The model accounts for the orthogonally polarized modes
(OPM) in the emission and has been used to argue that OPM occur simultaneously, 
not separately (MS1; McKinnon \& Stinebring 2000, hereafter MS2; McKinnon 
2002). The model also accounts for deviations from mode orthogonality and 
has been used to show that the modal connecting bridge in PSR B2016+28 at 
1404 MHz is a transition between modes of nonorthogonal polarization 
(McKinnon 2003a), thereby resolving a possible discrepancy in the 
interpretation of the pulsar's viewing geometry (SCRWB). Additionally,
the model has been used to show that the large fluctuations in fractional 
circular polarization observed in pulsar radio emission (SCRWB) may be 
attributed to the heavy modulation of the mode intensities and the small 
degree of circular polarization intrinsic to the modes (McKinnon 2002). 

  The objective of this paper is to derive and apply the analytical tools 
necessary to determine the origin of the excess PA dispersion in pulsar 
radio emission. An eigenvalue analysis of radio polarization measurements 
that is based upon the three-dimensional statistics of radio polarimetry 
(McKinnon 2003b) is presented in \S~\ref{sec:eigenval}. In the analysis, 
eigenvalues of polarization measurements are derived for different 
polarization models, such as fixed polarization, RPR, random fluctuations 
in PA, OPM, and nonorthogonal polarization modes. The eigenvalue analysis 
is applied to sensitive, well-calibrated polarization observations of the 
nearby pulsars PSR B1929+10 and PSR B2020+28 in \S~\ref{sec:apply}. The 
analysis indicates that the excess PA dispersion is caused by RPR in the 
received pulsar signal. In \S~\ref{sec:discuss}, the nature of polarization 
fluctuations in pulsar radio emission is clarified, the origin and 
consequences of RPR are discussed, and additional applications of the 
eigenvalue analysis are proposed. The conclusions of the analysis are 
summarized in \S~\ref{sec:conclude}.

\section{Eigenvalue Analysis of Radio Polarimetry}
\label{sec:eigenval}

  The Stokes parameters Q, U, and V completely describe the polarization 
of electromagnetic radiation. Fortunately for radio astronomers, all three 
Stokes parameters of a radio signal can be measured simultaneously, so the 
polarization state of the radiation can be completely determined with a 
single measurement (Radhakrishanan 1999; McKinnon 2003b). When multiple
measurements of the Stokes parameters are plotted in a three-dimensional 
Cartesian coordinate system, or Poincar\'e space in what follows, the 
shape of the resulting data point cluster generally resembles an 
ellipsoid. The location of the ellipsoid's centroid is given by the 
average values of the Stokes parameters. The dimensions and orientation 
of the ellipsoid are determined by polarization fluctuations intrinsic 
to the radio source and by the instrumental noise. It follows that one 
may determine what causes the polarization fluctuations provided the 
dimensions and orientation of the ellipsoid can be measured.

  The dimensions and orientation of the polarization ellipsoid can be 
measured with an eigenvalue analysis of the Stokes parameters. The
covariances of the three Stokes parameters form a 3x3 data matrix. The 
eigenvalues of this covariance matrix are measures of the ellipsoid's 
three dimensions. The eigenvectors associated with the eigenvalues are 
the three axes, or the orientation, of the ellipsoid. The dimensions of 
the ellipsoid are largely unaffected by whether the source polarization 
occurs in Q, U, V, or any combination of the three Stokes parameters. 
Therefore, the eigenvalue analysis and the conclusions drawn from it 
are generally independent of whether the radio source is linearly-, 
circularly-, or elliptically-polarized. In the following examples, 
eigenvalues are derived for different types of mechanisms that create 
polarization fluctuations.

\begin{figure}
\plotone{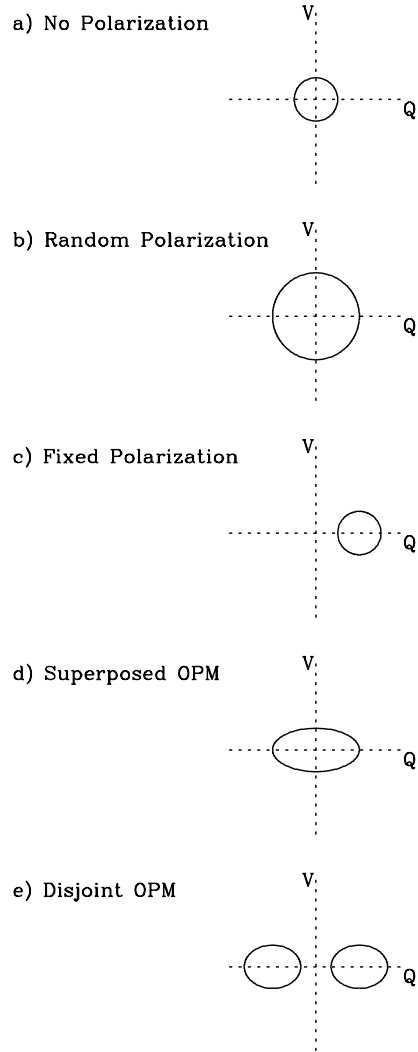}
\caption{Shapes of Q-U-V data point clusters for different types of 
polarization. The shapes are shown as cross-sections of the cluster in
the Q-V plane of Poincar\'e space. The cluster is a sphere for no 
polarization (a), random polarization (b), and fixed polarization (c). 
The cluster resembles a prolate ellipsoid for superposed OPM (d). For 
disjoint OPM (e), the polarization data reside in two, diametrically-opposed 
ellipsoids. The signal in (c) and the OPM in (d) and (e) are linearly
polarized.} 
\label{fig:clusters}
\end{figure}

\subsection{Fixed Polarization}

  Let us consider a radio source with constant linear polarization, $\mu$, 
and assume that its polarization occurs in the Stokes parameter Q. When a 
radio telescope is pointed towards the source, the measured Stokes 
parameters are

\begin{equation}
{\rm Q} = \mu + X_{\rm N,Q},
\end{equation}
\begin{equation}
{\rm U} = X_{\rm N,U},
\end{equation}
\begin{equation}
{\rm V} = X_{\rm N,V},
\end{equation}
where $X_{\rm N}$ is a Gaussian random variable with zero mean and standard
deviation, $\sigma_N$, that represents the instrumental noise. When multiple
measurements of these Stokes parameters are plotted in Poincar\'e space, the 
resulting cluster of data points is spherical in shape. The sphere has a 
radius that is characterized by the magnitude of the instrumental noise, 
$\sigma_N$, and its centroid is offset from the coordinate system origin by 
a distance $\mu$ (Fig.~\ref{fig:clusters}c).

  The dimensions of the sphere can be determined through an eigenvalue 
analysis of the Stokes parameters. The matrix formed by the covariances 
of the Stokes parameters is
\begin{equation}
\bar C = \pmatrix{{\rm Cov(Q,Q)} & {\rm Cov(Q,U)} & {\rm Cov(Q,V)} \cr
             {\rm Cov(U,Q)} & {\rm Cov(U,U)} & {\rm Cov(U,V)} \cr
             {\rm Cov(V,Q)} & {\rm Cov(V,U)} & {\rm Cov(V,V)} \cr}
  = \sigma_N^2\pmatrix{1 & 0 & 0 \cr
                     0 & 1 & 0 \cr
                     0 & 0 & 1 \cr}.
\end{equation}
Regardless of radio source polarization, the elements of the covariance 
matrix are always real and the matrix is always symmetric about its diagonal 
because the Stokes parameters are always real numbers (i.e., the covariance 
matrix is always Hermetian). In this particular example, the off-diagonal 
elements of $\bar C$ are equal to zero because the instrumental noise is 
independent between Stokes parameters (MS1; MS2). Since $\bar C$ is symmetric 
about its diagonal and its off-diagonal elements are equal to zero, the 
matrix forms the eigenbasis of the data point cluster, and the elements on 
the matrix diagonal are its eigenvalues. The square root of each eigenvalue 
is equal to the magnitude of the instrumental noise, $\sigma_N$, which is
proportional to the size of the sphere's radius. The actual radius of the
sphere is about $3\sigma_N$ because the fluctuations in instrumental noise 
are Gaussian and, thus, approximately $99.73\%$ of the Q-U-V data points
reside within a sphere of this radius. Since the eigenvalues are equal to one 
another, the matrix is triply-degenerate, which means that the eigenvectors 
corresponding to the eigenvalues do not uniquely define the three axes of 
the data point cluster, as one would expect for a sphere. 

  The matrix formed by the second moments of the Stokes parameters is
\begin{equation}
\bar S  = \sigma_N^2\pmatrix{1+s^2 & 0 & 0 \cr
                              0 & 1 & 0 \cr
                              0 & 0 & 1 \cr}
\end{equation}
where $s=\mu/\sigma_N$ is the signal-to-noise ratio in the polarization
vector amplitude. The matrix $\bar S$ is also an eigenbasis. The principal 
eigenvector of $\bar S$ has an eigenvalue of $\sigma_N^2(1+s^2)$. It is 
aligned with the polarization vector of the radio source, which in this
example is the Stokes parameter Q.

  If the radio source emits RPR, the Q-U-V data point cluster retains its 
spherical shape, but is inflated by an amount equal to the random polarization 
fluctuations, $\sigma_P$ (Fig.~\ref{fig:clusters}b). The eigenvalues of the 
covariance matrix become $\tau_{11}=\tau_{22}=\tau_{33}=\sigma_N^2+\sigma_P^2$, 
and the sphere's radius is proportional to the square root of these eigenvalues.
If the random polarization fluctuations are also Gaussian, the measured standard
deviation in PA is
\begin{equation}
\sigma_\psi = {1\over{2\mu}}(\sigma_P^2+\sigma_N^2)^{1/2}.
\end{equation}

\subsection{Random Fluctuations in Position Angle}

  Stochastic Faraday rotation and random fluctuations in the orientation of a 
pulsar's magnetic field are possible explanations for the excess PA dispersion 
in the emission. Both processes can be modeled as random fluctuations in PA. 
For a linear polarization vector with fixed amplitude and an orientation
that randomly varies in the Q-U plane of Poincar\'e space (i.e. in azimuth 
only), the measured Stokes parameters are 

\begin{equation}
{\rm Q} = \mu\cos\phi  + X_{\rm N,Q},
\end{equation}
\begin{equation}
{\rm U} = \mu\sin\phi  + X_{\rm N,U},
\end{equation}
\begin{equation}
{\rm V} = X_{\rm N,V},
\end{equation}
where $\phi$ is a random variable that accounts for the vector's fluctuations
in azimuth. If the fluctuations in vector azimuth are Gaussian with zero mean 
and a standard deviation of $\sigma_\phi$, the measured polarization occurs 
primarily in Stokes Q, and the eigenvalues of the covariance matrix are
\begin{equation}
\tau_{11}= {\mu^2\over{2}}[1-\exp{(-\sigma_\phi^2)}]^2 + \sigma_N^2,
\end{equation}
\begin{equation}
\tau_{22}= {\mu^2\over{2}}[1-\exp{(-2\sigma_\phi^2)}] + \sigma_N^2,
\end{equation}
\begin{equation}
\tau_{33} = \sigma_N^2.
\end{equation}
The shape of the Q-U-V data point cluster formed by Gaussian fluctuations in 
PA is a banana-shaped, arcing ellipsoid. The centroid of the arcing ellipsoid 
resides in the Q-U plane at a distance $\mu$ from the origin of Poincar\'e 
space. When the PA fluctuations are small ($\sigma_\phi \ll 1$), 
$\tau_{11}\simeq\tau_{33}$, and the cluster is a prolate ellipsoid (i.e., 
an ellipsoid that is rotationally symmetric about its major axis). The 
ellipsoid's major axis is aligned with the Stokes parameter U because the 
polarization fluctuations are largest in U (i.e., 
$\tau_{22} > \tau_{11} > \tau_{33}$). Therefore, the major axis of the 
ellipsoid is perpendicular to the orientation of the radiation's polarization 
vector. Since polarization position angle is related to polarization vector 
azimuth by $\psi=\phi/2$, the standard deviation in the measured PA is 
\begin{equation}
\sigma_\psi = {1\over{2}}(s^{-2}+\sigma_\phi^2)^{1/2}.
\end{equation}

\subsection{Orthogonal Modes of Polarization}
\label{sec:opm}

  In the statistical model of pulsar polarization proposed by
McKinnon \& Stinebring (MS1; MS2), the observed polarization is 
determined by the simultaneous interaction of two, orthogonally 
polarized modes with randomly varying polarization amplitudes. 
Since the modes are orthogonal, their polarization vectors lie on 
the same diagonal in Poincar\'e space, regardless of the degree 
of their linear or circular polarization. The statistical model 
designates the amplitudes of the mode polarizations by the random 
variables $X_1$ and $X_2$, which have means $\mu_1$ and $\mu_2$ 
and standard deviations $\sigma_1$ and $\sigma_2$, respectively. 
Assuming for this particular example that the modes are linearly 
polarized and that their polarization vectors lie along the diagonal 
aligned with the Stokes parameter Q, the measured Stokes parameters 
are
\begin{equation}
{\rm Q} = X_1 - X_2 + X_{\rm N,Q},
\label{eqn:Qopm}
\end{equation}
\begin{equation}
{\rm U} = X_{\rm N,U},
\label{eqn:Uopm}
\end{equation}
\begin{equation}
{\rm V} = X_{\rm N,V}.
\label{eqn:Vopm}
\end{equation}
If $X_1$ and $X_2$ are independent random variables, the covariance 
matrix computed from these Stokes parameters is
\begin{equation}
\bar C=\sigma_N^2\pmatrix{1 + \rho^2 & 0 & 0 \cr
                     0 & 1 & 0 \cr
                     0 & 0 & 1 \cr},
\end{equation}
where $\rho=(\sigma_1^2+\sigma_2^2)^{1/2}/\sigma_N$ is the magnitude of the 
intrinsic polarization fluctuations relative to the instrumental noise. The 
shape of the Q-U-V data point cluster formed by superposed OPM resembles a 
prolate ellipsoid\footnote{Even if the mode polarization amplitudes are not 
independent, the cluster is still a prolate ellipsoid but with a principal 
eigenvalue of 
$\tau_{11}=\sigma_1^2 + \sigma_2^2 - 2r_{12}\sigma_1\sigma_2 + \sigma_N^2$,
where $r_{12}$ is the correlation coefficient of the polarization amplitudes 
(MS1).} (see Fig.~\ref{fig:clusters}d). The centroid of the ellipsoid is 
offset from the origin of Poincar\'e space by a distance $\mu=|\mu_1-\mu_2|$. 
The dimensions of the ellipsoid are proportional to the square root of the 
eigenvalues. The size of the ellipsoid's major axis scales as 
$a_1=\sqrt\tau_{11}=\sigma_N(1+\rho^2)^{1/2}$. The dimensions of the ellipsoid's 
minor axes are scaled by the instrumental noise, $a_2=a_3=\sigma_N$. The 
orientations of the ellipsoid's minor axes are not uniqely determined because 
the two minor eigenvalues are equal to each other. Unlike the case of random 
PA fluctuations considered above, the major axis of the OPM ellipsoid is parallel 
to the radiation's polarization vector. Thus, the OPM model predicts that the 
principal eigenvectors determined from the covariances and second moments of 
the Stokes parameters will be aligned. 

  The axial ratio of the ellipsoid can be related to the modulation index 
of the emission. For example, if the modes are completely polarized, the 
polarization fluctuations are given by $\rho=\beta s$, where $\beta$ and $s$ 
are the modulation index and signal to noise ratio, respectively, of the total 
intensity (McKinnon 2002). The ellipsoid axial ratio is
\begin{equation}
a_1/a_3=(1+\rho^2)^{1/2}=(1 + \beta^2s^2)^{1/2},
\end{equation}
which becomes larger as the emission becomes more heavily modulated and as 
the signal to noise ratio of the observation improves.  At the other extreme
when the signal to noise ratio is poor or the polarization does not fluctuate
(i.e. $\rho=\beta=0$), the axial ratio approaches unity, and the cluster is 
a spheroid.

  The detailed shape and dimensions of an OPM Q-U-V cluster can be affected 
by the character of the fluctuations in $X_1$ and $X_2$. Simple numerical 
simulations of equations~\ref{eqn:Qopm} through~\ref{eqn:Vopm} indicate that 
the cluster resembles a prolate ellipsoid whenever $\sigma_1\simeq\sigma_2$, 
regardless of the character of the fluctuations. The simulations also show
that Gaussian fluctuations in $X_1$ and $X_2$ produce a Q-U-V cluster that is 
a prolate ellipsoid, even if $\sigma_1$ differs from $\sigma_2$. When $X_1$ 
and $X_2$ are exponential random variables with $\sigma_1\gg\sigma_2$, the 
ellipsoid is still rotationally symmetric about its major axis, but the 
ellipsoid is stretched along its major axis in the direction of the dominant 
mode, $X_1$. This is to be expected because the polarization fluctuations 
would be largely determined by the dominant mode. The dimensions of the 
cluster's minor axes extend about $\pm 3a_2$ and $\pm 3a_3$ from the cluster
centroid because the size of the minor axes are set by the Gaussian 
instrumental noise. The size of the cluster's major axis extends about 
$\pm 3a_1$ from the centroid if the mode polarizations are Gaussian random 
variables and can be larger if they are exponential random variables.

  If OPM are disjoint (i.e. occur separately) instead of superposed, the 
Q-U-V data points will reside in two ellipsoids that are diametrically 
opposed in Poincar\'e space (Fig.~\ref{fig:clusters}e). The size of the
ellipsoids' minor axes are set by the instrumental noise. The size of the 
ellipsoids' major axes are determined by the instrumental noise and the
polarization fluctuations of the modes.

\subsection{Nonorthogonal Polarization Modes}
\label{sec:npm}

  Nonorthogonal polarization modes do not occur along the same diagonal in 
Poincar\'e space. However, the two vectors representing the modes still 
define a plane, and the orientation of this plane depends upon the 
elliptical polarization of the modes. Here, and without loss of generality, 
the modes are assumed to be linearly polarized. If the deviation from 
orthogonality is given by the angle $\theta$, incorporating the 
nonorthogonality of the modes in the model described above (McKinnon 2003a)
gives the following expressions for the measured Stokes parameters.
\begin{equation}
{\rm Q} = X_1 - X_2\cos\theta + X_{\rm N,Q},
\end{equation}
\begin{equation}
{\rm U} = X_2\sin\theta + X_{\rm N, U},
\end{equation}
\begin{equation}
{\rm V} = X_{\rm N,V}.
\end{equation}
If the mode polarizations are independent (i.e., $\rm{Cov}(X_1,X_2)=0$), 
the covariance matrix for the nonorthogonal modes is 
\begin{equation}
\bar C=\pmatrix{\sigma_1^2+\sigma_2^2\cos^2\theta + \sigma_N^2 & 
   -\sigma_2^2\cos\theta\sin\theta & 0 \cr
   -\sigma_2^2\cos\theta\sin\theta & \sigma_2^2\sin^2\theta +\sigma_N^2 & 0 \cr
   0 & 0 & \sigma_N^2 \cr}.
\label{eqn:nonortho}
\end{equation}
The eigenvalues of the covariance matrix in equation~\ref{eqn:nonortho} are
\begin{equation}
\tau_{11} = {1\over{2}}\Biggl[(\sigma_1^2 + \sigma_2^2 + 2\sigma_N^2)
         + \sqrt{\sigma_1^4 + \sigma_2^4 + 2\sigma_1^2\sigma_2^2\cos(2\theta)}
           \Biggr],
\end{equation}
\begin{equation}
\tau_{22} = {1\over{2}}\Biggl[(\sigma_1^2 + \sigma_2^2 + 2\sigma_N^2)
         - \sqrt{\sigma_1^4 + \sigma_2^4 + 2\sigma_1^2\sigma_2^2\cos(2\theta)}
           \Biggr],
\end{equation}
\begin{equation}
\tau_{33} = \sigma_N^2.
\end{equation}
The shape of the Q-U-V data point cluster formed by nonorthogonal polarization 
modes also resembles an arcing ellipsoid because the nonorthogonal mode breaks 
the symmetry of the cluster. 

\section{APPLICATION OF EIGENVALUE ANALYSIS}
\label{sec:apply}

\begin{figure}
\plotone{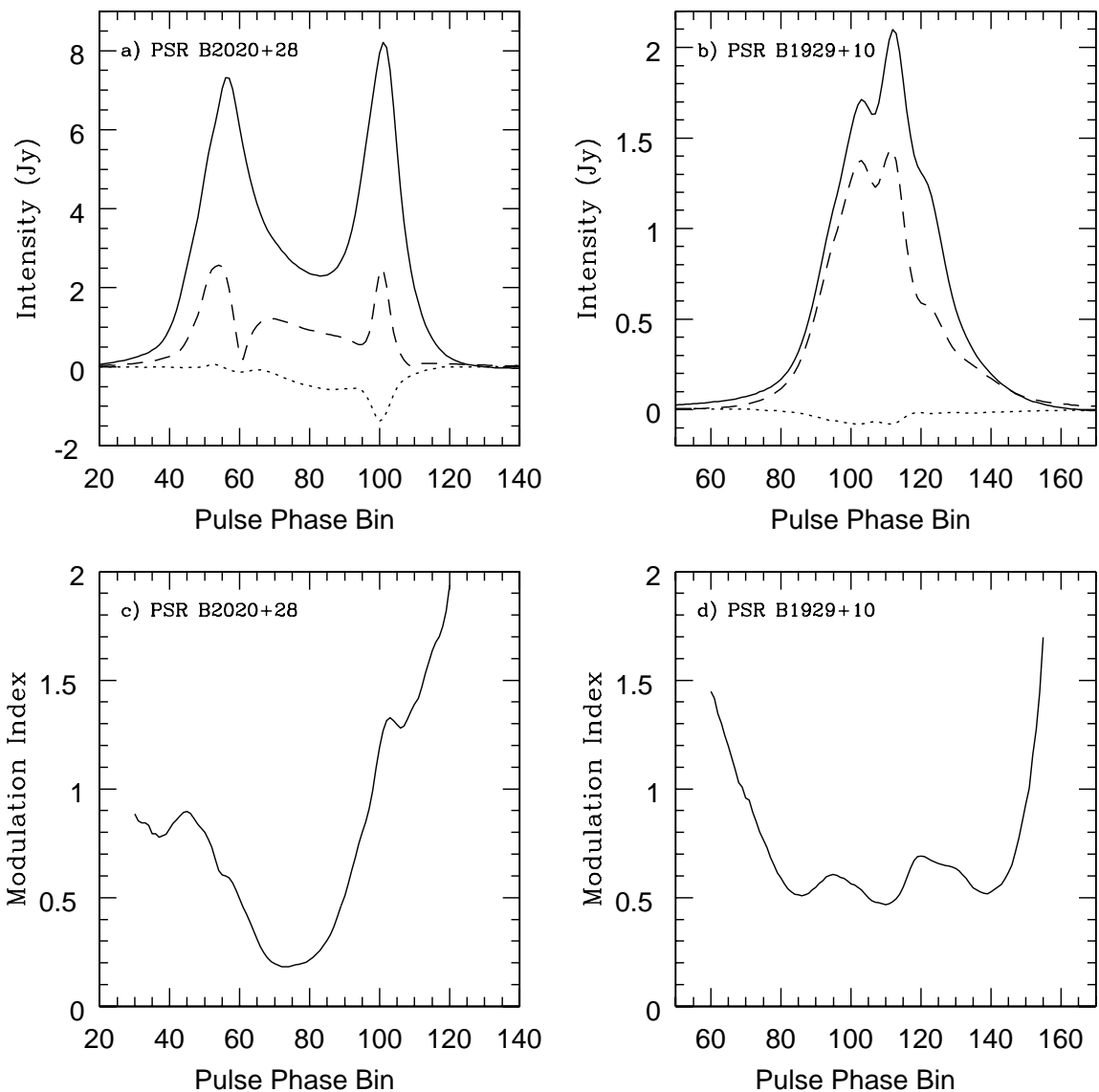}
\caption{Pulse profile and intensity modulation indices of PSR B2020+28 and 
PSR B1929+10 at 1404 MHz. The average profiles of the two pulsars are shown 
in the top panels. Total intensity, linear polarization, and circular 
polarization are denoted by the solid, dashed, and dotted lines, respectively. 
The bottom panels show how the total intensity modulation index varies across 
the pulse of each pulsar.}
\label{fig:profile}
\end{figure}

  The eigenvalue analysis developed in \S~\ref{sec:eigenval} was applied to 
sensitive, well-calibrated, polarization observations of PSR B2020+28 and 
PSR B1929+10 to determine the origin of the excess PA dispersion. The 
observations were made by SCRWB with the Arecibo radio telescope at a 
frequency of 1404 MHz. At the time the observations were made, the 
telescope's nominal system temperature was 40 K, and its forward gain 
was 8 K/Jy. Data from the same observations were used in the analyses 
documented in MS1, MS2, and McKinnon (2003a). SCRWB's superb polarization 
calibration is essential to the success of the eigenvalue analysis because 
improper or inadequate calibration will alter the dimensions of the Q-U-V 
data point cluster, and thus adversely influence the conclusions drawn from 
the analysis. To insure that the observed polarization fluctuations were 
intrinsic to the pulsar, and not due to slowly-varying signal amplification 
caused by diffractive interstellar scintillation (MS1), a subset of the 
overall time series where the running mean intensity did not vary was 
selected for further analysis. The time series thus selected contained 800 
pulses for PSR B2020+28 and 4000 pulses for PSR B1929+10. The pulse profiles 
and modulation indices computed from the selected time series are shown in 
Figure~\ref{fig:profile}.

  The eigenvalue analysis was performed as follows. For both pulsars, a 
covariance matrix of the Stokes parameters Q, U, and V was computed at each 
pulse phase bin. The covariances were then used to calculate the eigenvalues 
and eigenvectors of the matrix. As illustrated by the examples in 
\S~\ref{sec:eigenval}, the dimensions of the Q-U-V ellipsoid at each phase 
bin were estimated by taking the square root of the eigenvalues. Examples of 
Q-U-V ellipsoids for each pulsar are shown in Figures~\ref{fig:ell2020} 
and~\ref{fig:ell1929}. In each panel of both figures, every data point has 
been rotated by the PA at the relevant phase bin so that the major axis of 
the polarization ellipsoid occurs in the Q-V plane of Poincar\'e space. The 
rotation eliminates projection effects in the data presentation. The bottom 
panels of Figure~\ref{fig:tauratio} show how the ellipsoid axial ratios vary 
across the pulse of each pulsar. The ratio of the ellipsoid's largest 
dimension to its smallest dimension, $a_1/a_3$, is shown by the solid line 
in the bottom panels of the figure. The ratio of the ellipsoid's two smaller 
dimensions, $a_2/a_3$, is shown by the dotted line in the same panels.

  Eigenvalues and eigenvectors were also computed from the second moments of 
the Stokes parameters to insure that this new data analysis method reproduced 
the results obtained from more traditional methods. As predicted by the 
eigenvalue analysis, the PAs of polarization vectors given by the second 
moment principal eigenvectors were very similar to the PAs computed in the 
traditional way, 
\begin{equation}
\psi={1\over{2}}\arctan\Biggl({U\over{Q}}\Biggr).
\end{equation}
Furthermore, the vector colatitude given by the second moment principal 
eigenvector was very similar to the colatitude traditionally calculated 
from 
\begin{equation}
\theta=\arccos\Biggl[{V\over{(Q^2+U^2+V^2)^{1/2}}}\Biggr].
\end{equation}

\subsection{PSR B2020+28}
\label{sec:2020}

\begin{figure}
\plotone{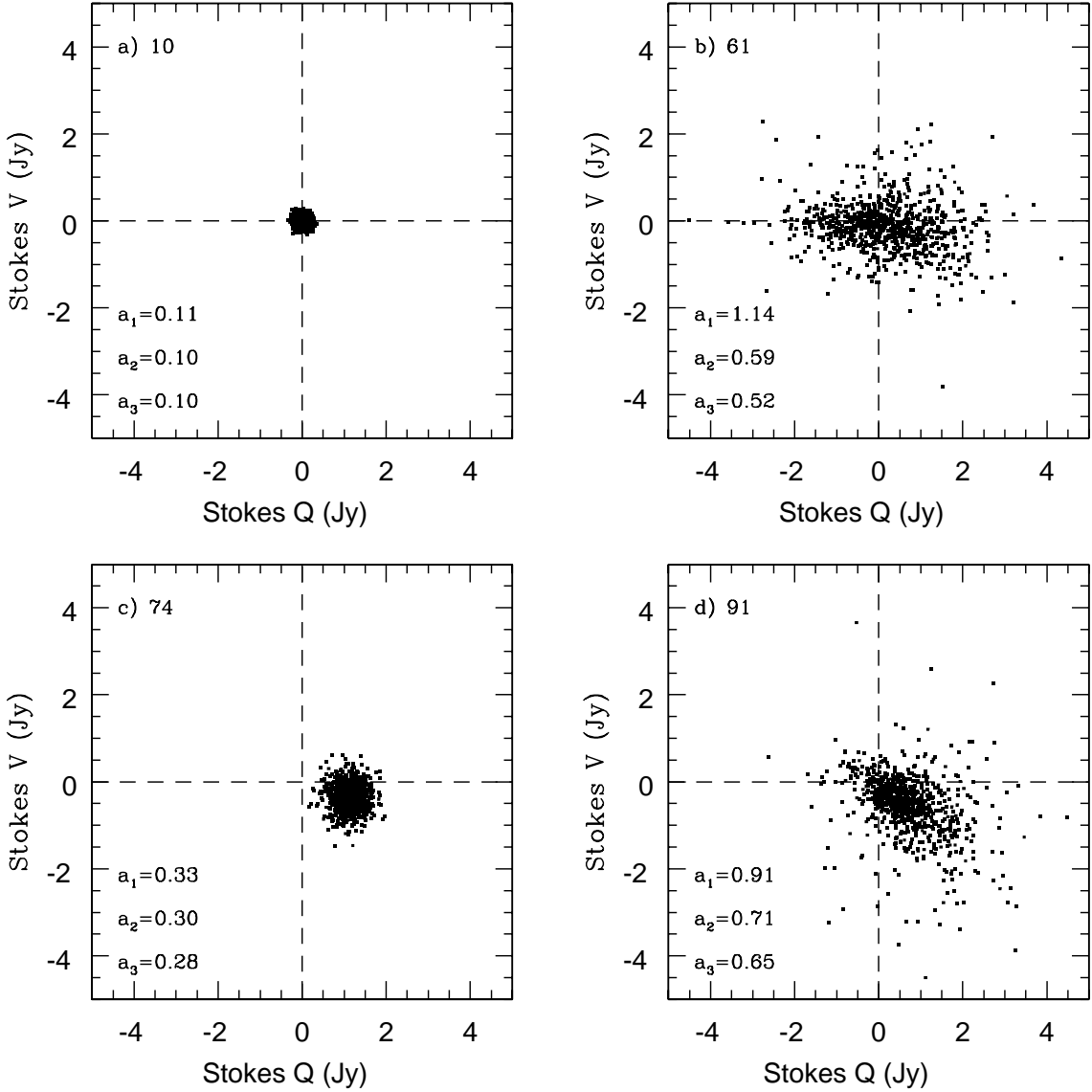}
\caption{Q-U-V data point clusters at different locations in the pulse of
PSR B2020+28. The pulse phase bin is denoted in the upper left corner of 
each panel. The quantities $a_1, a_2,$ and $a_3$ in the bottom left corner 
of each panel are the square roots of the cluster eigenvalues, in units of
Jy, and are proportional to the size of the cluster's three dimensions. For 
display purposes, the data points have been rotated so that the major axis 
of the cluster occurs in the Q-V plane of Poincar\'e space. OPM occur at 
bins 61 and 91, but only one polarization mode occurs at bin 74. The overall 
size of each cluster is much larger than the off-pulse noise at bin 10, 
suggesting that randomly polarized radiation accompanies the radio emission.}
\label{fig:ell2020}
\end{figure}

  Figure~\ref{fig:ell2020} shows four examples of Q-U-V data point clusters 
in PSR B2020+28 and provides the clue that explains the excess PA dispersion 
in the pulsar. Data recorded off the pulse are shown by the cluster at phase 
bin 10. The cluster has the spheroidal shape and size expected for instrumental 
noise. The cluster's values of $a_1, a_2,$ and $a_3$ are equal to each other
and are equal to the noise calculated off the pulse, $\sigma_N=0.1$ Jy. 
These values suggest that the cluster's three dimensions are equal and,
consequently, that the cluster is spheroidal in shape. All the Q-U-V data 
points at bin 10 reside within a sphere of radius, $r=3\sigma_N$. But contrary 
to most examples given in \S~\ref{sec:eigenval}, the smallest dimension of 
every other cluster shown in Figure~\ref{fig:ell2020}, as indicated by the 
values of $a_3$, is much greater than what is expected from the instrumental 
noise. 

  At phase bin 74, the cluster is a spheroid that is three times larger than 
what is expected from the off-pulse noise. The fact that the cluster at bin 
74 is a spheroid, and not an arcing ellipsoid, suggests that the excess PA 
dispersion at this pulse location is not caused by random fluctuations in PA 
(e.g. stochastic Faraday rotation or fluctuations in the orientation of 
magnetic field lines). Since the data point cluster is not a prolate 
ellipsoid, the data do not support the assumption of MS1 and MS2 that both 
polarization modes occur at bin 74 of PSR B2020+28. The large size of the 
spheroid, as well as the large values of $a_3$ at bins 61 and 91, suggests 
that the excess PA dispersion is caused by a mechanism that isotropically 
inflates the clusters. The only mechanism considered in \S~\ref{sec:eigenval}
that can isotropically inflate the cluster is RPR in the received signal.

\begin{figure}
\plotone{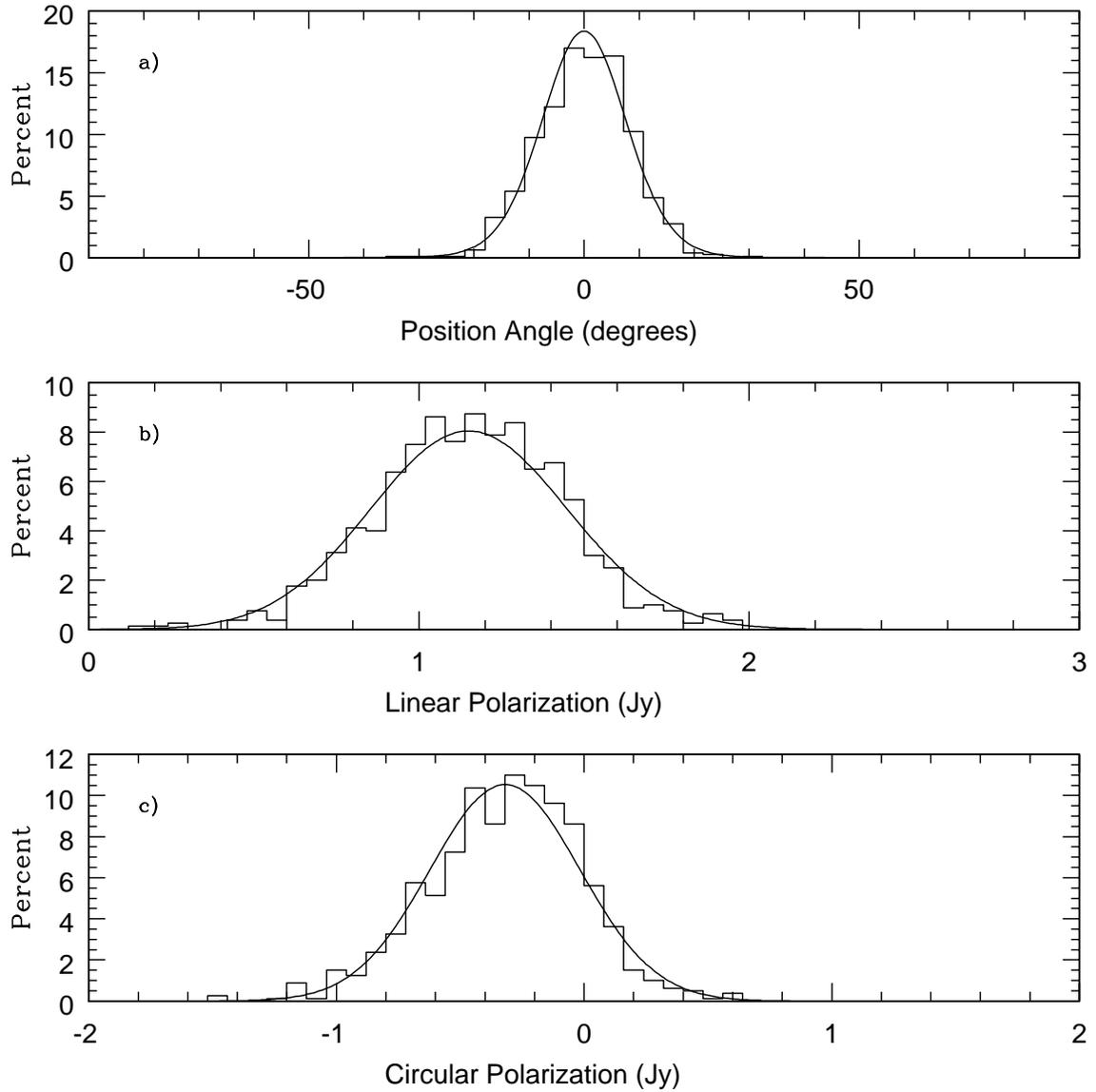}
\caption{Distributions of polarization position angle (a), linear polarization
(b), and circular polarization (c) at pulse phase bin 74 of PSR B2020+28. The 
histograms were constructed from measured polarization values. The solid, 
continuous lines in each panel are the theoretical distributions expected for 
a fixed polarization vector and Gaussian noise.}
\label{fig:PA74}
\end{figure}

  Another intriguing and noteworthy feature of the data at bin 74 is that 
the fluctuations in total intensity and polarization are Gaussian-like (see 
Fig.~\ref{fig:PA74} and Figs. 4 \& 7 of MS1). This is in stark contrast 
to intensity fluctuations in other pulsars, which can be lognormal, power law, 
and chi-squared or gamma (e.g., Cairns, Johnston, \& Das 2003; Cairns et al. 
2003; MS1; Cordes 1976a, 1976b). The emission's extremely low modulation index 
near the center of the pulse is also indicative of Gaussian intensity fluctuations. 
Since total intensity must be non-negative, Gaussian intensity fluctuations 
require the mean intensity, $\mu_I$, to be at least a factor of three to five 
times greater than the intensity standard deviation, $\sigma_I$. Therefore, 
the modulation index of Gaussian fluctuations must satisfy 
$\beta=\sigma_I/\mu_I < 0.2-0.3$. Interestingly, at phase bins 65 to 85 where 
the modulation index is $\beta<0.3$, the Q-U-V data point clusters also have 
spheroidal shapes, similar to that of bin 74. Thus, the polarization properties 
at bins 65 through 85 are similar to those at bin 74.

  The simplest explanation for the Q-U-V data point cluster at bin 74 is a 
combination of a fixed polarization vector, RPR, and instrumental noise. 
If this hypothesis is correct, the observed Gaussian fluctuations in 
polarization would be due to a combination of RPR and instrumental noise 
because the polarization vector is assumed to have a fixed amplitude. 
Furthermore, the distributions of PA, linear polarization, and circular 
polarization should all be broadened by the fluctuations due to RPR. To test 
this hypothesis, histograms of linear polarization, circular polarization, 
and PA constructed from the bin 74 polarization data were compared to the 
theoretical distributions expected for a fixed polarization vector in Gaussian 
noise (McKinnon 2002), where the noise is due to a combination of RPR and 
instrumental noise.
\begin{equation}
f_\psi(\psi) = {1\over{\pi}}
   \exp{\Biggl(-{\mu_{\rm L}^2\over{2\sigma^2}}\Biggr)}
   \Biggl\{1 + \sqrt{{\pi\over{2}}}{\mu_{\rm L}\over{\sigma}}\cos(2\psi)
    \exp{\Biggl({\mu_{\rm L}^2\over{2\sigma^2}}\cos^2(2\psi)\Biggr)}
    \Biggl[1+{\rm erf}\Biggl({\mu_{\rm L}\over{\sigma}}
    {\cos(2\psi)\over{\sqrt{2}}}\Biggr)\Biggr]\Biggr\}
\label{eqn:phidist}
\end{equation}
\begin{equation}
f_{\rm L}(x) = {x\over{\sigma^2}}
               \exp{\Biggl[-{(\mu_{\rm L}^2 + x^2)\over{2\sigma^2}}\Biggr]}
               I_0\Biggl({x\mu_{\rm L}\over {\sigma^2}}\Biggr)
\label{eqn:Ldist}
\end{equation}
\begin{equation}
f_{\rm V}(x) = {1\over{\sigma\sqrt{2\pi}}}
               \exp{\Biggl[-{(x - \mu_{\rm V})^2\over{2\sigma^2}}\Biggr]}
\label{eqn:Vdist}
\end{equation}
The parameters used to construct the theoretical distributions were 
obtained from the experimental data. The mean circular polarization was 
$\mu_V= -0.32$ Jy, the mean linear polarization was $\mu_L=(\mu_Q^2+
\mu_U^2)^{1/2}=1.11$ Jy, and the standard deviation of the Gaussian noise 
was taken from the bin 74 data point cluster as $\sigma=0.30$ Jy. As shown 
in Figure~\ref{fig:PA74}, the theoretical distributions compare extremely 
well with the measured histograms. The comparison of the PA distributions 
is much more favorable than the result obtained by MS1\footnote{The 
histograms in Figs. 7b and 7d of MS1 utilize smaller bin sizes than those of 
Fig.~\ref{fig:PA74}. Therefore, the histograms appear to be slightly different, 
although the same data were used in the two figures.} where identical data 
were used but the standard deviation of the Gaussian noise was assumed to 
be $\sigma=0.1$ Jy, instead of 0.3 Jy, in the calculation of the theoretical 
distribution. In summary, the favorable comparison between the theoretical 
and measured distributions in Figure~\ref{fig:PA74} confirms a simple 
interpretation of the emission's polarization properties at bin 74. The
Gaussian fluctuations in polarization are due to RPR and instrumental noise.
These fluctuations broaden the PA distribution as well as the distributions 
of linear and circular polarization. Therefore, the excess dispersion in PA 
is due solely to RPR. Another component of the emission has a polarization
vector with fixed amplitude and orientation. Consequently, only one
polarization mode with constant polarization occurs at bin 74.

  Given this interpretation, one may also say something about the fluctuations 
in total intensity at bin 74. Since the mode's polarization is constant, the 
mode's intensity must also be constant because it is highly unlikely that the 
mode's intensity can randomly fluctuate while its polarization remains constant, 
particularly when the mode is completely polarized as assumed in MS1 and MS2. 
Therefore, the observed fluctuations in total intensity at bin 74 cannot be 
attributed to the polarization mode. They must be due to RPR.

  Gaussian fluctuations and random polarization are properties of instrumental 
noise. However, the inflation of the Q-U-V data point clusters cannot be 
attributed to on-pulse instrumental noise. At bin 74, the average total 
intensity is 2.7 Jy, or an antenna temperature of 21.6 K. This change in 
antenna temperature would increase the noise by no more than a factor of 1.5 
(see eqn. (B4) of MS2), far short of the factor of three required to explain 
the size of the cluster. Furthermore, different locations within the pulse of 
PSR B2020+28 have similar total intensities, but the smallest dimensions of 
the clusters at these locations are very different (Fig.~\ref{fig:tauratio}), 
suggesting that the mechanism that inflates the cluster is not instrumental 
in origin. Also, the inflation of the clusters cannot be attributed to the 
correlation of on-pulse instrumental noise between Stokes parameters. 
Correlated instrumental noise will stretch a cluster in one dimension, 
instead of inflating it, because correlated data points will tend to form
a straight line in Poincar\'e space. For example, in the extreme case when 
the instrumental noise is completely correlated between the Stokes parameters, 
all elements of the covariance matrix will be $\sigma_N^2$. The matrix 
eigenvalues will be $\tau_{11}= 3\sigma_N^2, \tau_{22}=\tau_{33}=0$, which 
indicate that the data point cluster is one-dimensional. 

  Superposed OPM definitely occur at bins 61 and 91 of PSR B2020+28 because the 
data point clusters at these locations are elongated along the line that passes 
through the cluster centroid and the Q-V origin (the diagonal in Poincar\'e space). 
The simple fact that the cluster at each bin resembles a single, prolate ellipsoid 
and not two distinct ellipsoids suggests that OPM in this pulsar are superposed, 
not disjoint. At bin 61, the modes are primarily linearly polarized because the 
ellipsoid's major axis does not tilt much from $V=0$. If the polarization data 
at bin 61 were plotted in a PA histogram, the modes would be shown to occur with 
nearly equal frequency (see Fig. 5 of MS1) because the ellipsoid centroid roughly 
coincides with the Q-V origin and the number of data points on either side of 
$Q=0$ is about the same. At bin 91, the modes have a significant degree of 
circular polarization, as indicated by the ellipsoid's tilt. One mode would 
occur much more frequently than the other in a PA histogram because most data 
points have $Q>0$. RPR also accompanies the modes at these pulse locations 
because the two smaller ellipsoid dimensions indicated by $a_2$ and $a_3$ are 
approximately equal but are much larger than the instrumental noise.

  Figure~\ref{fig:tauratio} indicates where OPM occur in the pulse of PSR 
B2020+28 and shows how the magnitude of RPR may vary across it. The RPR 
magnitude, as estimated from the value of $a_3$ for each cluster and the 
off-pulse instrumental noise, 
\begin{equation}
\sigma_P=(a_3^2-\sigma_N^2)^{1/2},
\label{eqn:rpr}
\end{equation}
is shown in Figure~\ref{fig:tauratio}a. The magnitude of RPR is small in 
comparison to the total intensity at every location within the pulse. In 
Figure~\ref{fig:tauratio}c, the large ratios of the ellipsoids' major and 
minor axes, $a_1/a_3$, indicate where OPM occur, primarily in the peaks and 
wings of the pulse. The intensity modulation index is large 
(Fig.~\ref{fig:profile}c) at the same locations. The ratio 
of the ellipsoids' smaller dimensions is $a_2/a_3\simeq 1$ everywhere in the 
pulse, indicating that the ellipsoids are generally rotationally symmetric about 
their major axes as predicted by the OPM statistical model. If the small 
deviations from axial symmetry are real, they may be due to a slight 
nonorthogonality in the pulsar's polarization modes (McKinnon 2003a). The axial 
ratios and the modulation indices are small near the pulse center where it is 
possible that only one polarization mode with fixed polarization occurs. 

\subsection{PSR B1929+10}
  
\begin{figure}
\plotone{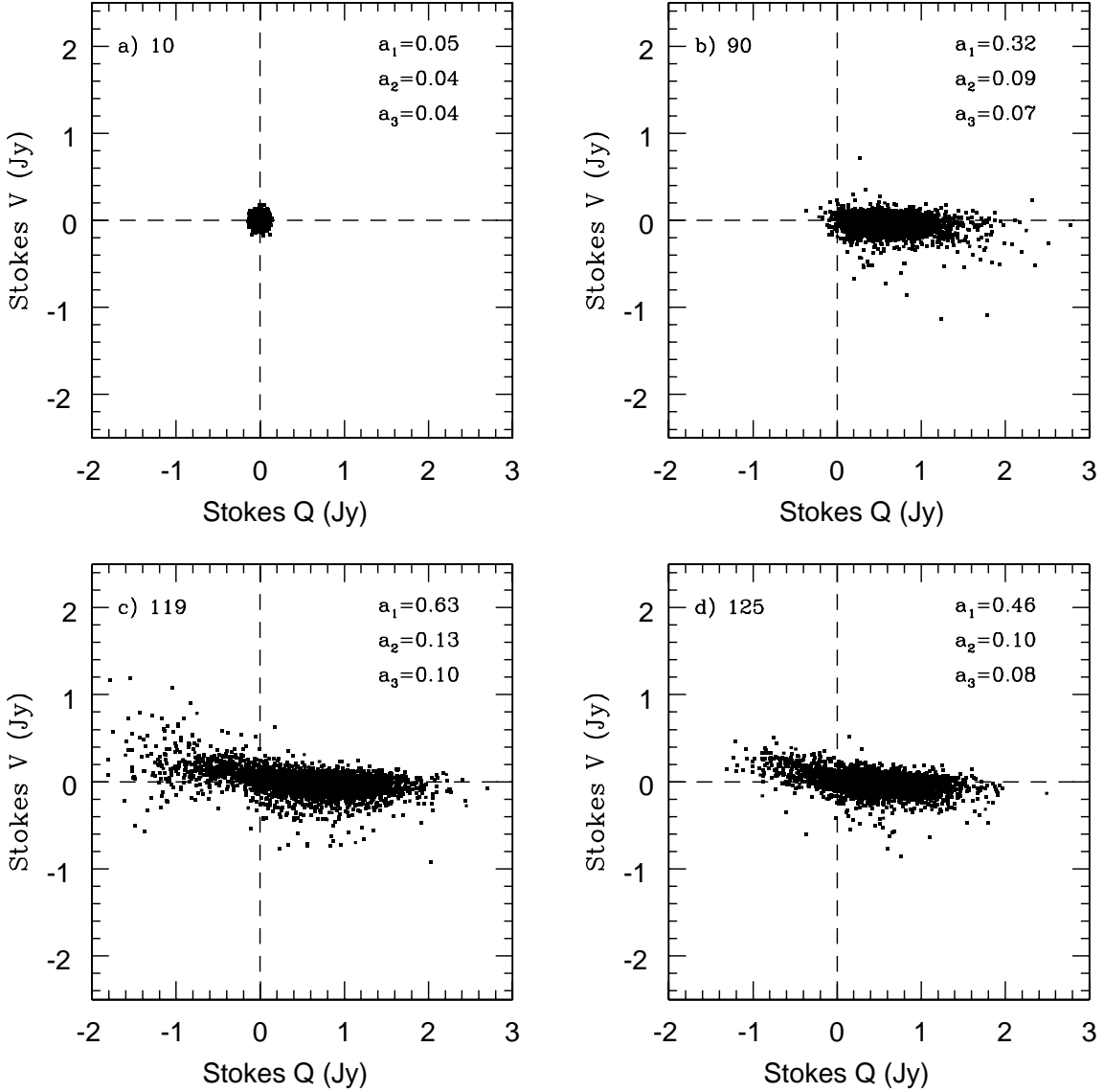}
\caption{Q-U-V data point clusters at different locations in the pulse of
PSR B1929+10. The pulse phase bin is denoted in the upper left corner of 
each panel, and the square roots of the cluster eigenvalues are listed in 
the upper right corner. The data points have been rotated so that the major 
axis of the cluster occurs in the Q-V plane of Poincar\'e space. OPM occur 
at each bin where the cluster is highly elongated. The modes at bins 119 
and 125 are not truly orthogonal because not all data points reside near 
the same diagonal. Similar to PSR B2020+28, the overall size of each cluster 
is larger than the off-pulse noise (bin 10).}
\label{fig:ell1929}
\end{figure}

  Figure~\ref{fig:ell1929} shows examples of Q-U-V data point clusters in PSR 
B1929+10. Again, the cluster representing the data recorded off the pulse has 
the spheroidal shape and size expected for the instrumental noise. The magnitude 
of the noise is smaller than that for PSR B2020+28 primarily because a larger 
bandwidth was used for the PSR B1929+10 observation. The smallest dimension 
of each ellipsoid is about a factor of two larger than expected from the 
instrumental noise, suggesting that the signal received from the pulsar contains 
RPR. 

  The clusters at phase bins 90, 119, and 125 are highly elongated, single 
ellipsoids, indicating the presence of superposed polarization modes. Since 
the data at each bin do not reside in two distinct clusters, the observations 
do not support the hypothesis of disjoint polarization modes. The superposed 
modes are primarily linearly polarized because the major axis of each ellipsoid 
is nearly aligned with $V=0$. Although both polarization modes occur at bin 90, 
as indicated by its highly elongated Q-U-V cluster, only one mode will appear 
in a PA histogram computed from these data because the vast majority of data 
points have $Q>0$. This simply means that one polarization mode is much stronger 
than the other (i.e., $\mu_1\gg\mu_2$). At phase bins 119 and 125, both 
polarization modes will appear in PA histograms because data points in these 
Q-U-V clusters reside on either side of $Q=0$. In both cases, one mode will 
occur more frequently than the other because more data points have $Q>0$ than 
$Q<0$. Figure 5 of McKinnon (2003a) shows the PA histogram for bin 119. The 
polarization modes at phase bins 119 and 125 are not precisely orthogonal 
because not all the data points in the Q-U-V clusters reside near a diagonal 
in Poincar\'e space. These clusters are consistent with what is expected for 
superposed, nonorthogonal polarization modes (see \S~\ref{sec:npm} above and 
McKinnon 2003a). Historically, cases of nonorthogonal polarization modes have 
been documented with PA histograms (Backer \& Rankin 1980; SCRWB; McKinnon 
2003a), which give no indication that the nonorthogonality may also occur in 
circular polarization. When the Q-U-V data are reported as polarization 
ellipsoids, the nonorthogonality in circular polarization is readily apparent. 

\begin{figure}
\plotone{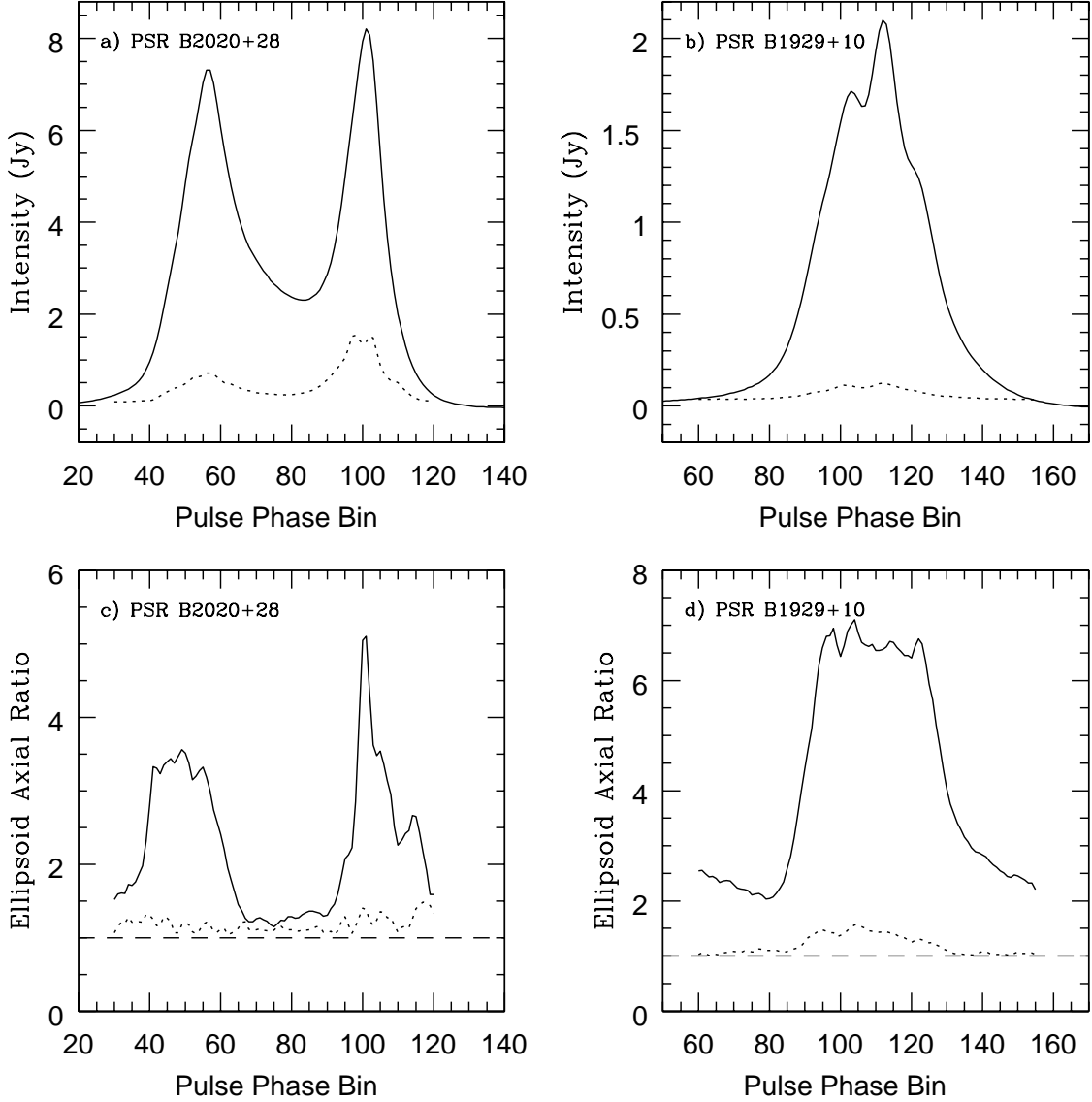}
\caption{Randomly polarized radiation (RPR) and the relative dimensions of the 
polarization ellipsoids in PSR B2020+28 and PSR B1929+10. The top panels show 
the total intensity (solid line) and the magnitude of RPR (dotted line) for 
each pulsar. The bottom panels show the relative dimensions of the polarization
ellipsoids for each pulsar. The solid line is the ratio of the largest to the 
smallest ellipsoid dimension, and the dotted line is the ratio of the two 
smaller ellipsoid dimensions. The ratios indicate that the ellipsoids are 
generally prolate, a result that is consistent with the statistical model of 
superposed OPM.}
\label{fig:tauratio}
\end{figure}

  Figure~\ref{fig:tauratio}b shows how the magnitude of RPR, as estimated 
from equation~\ref{eqn:rpr}, varies across the pulse of PSR B1929+10. As 
with PSR B2020+28, the RPR in PSR B1929+10 cannot be attributed to on-pulse 
instrumental noise because the total intensity of the pulsar cannot increase 
the noise by the factor required by the smallest dimensions of the polarization 
ellipsoids. The magnitude of RPR in PSR B1929+10 is much less than that in PSR 
B2020+28.

  The shapes of the polarization ellipsoids in the pulse of PSR B1929+10 
are consistent with the statistical model of superposed polarization modes. 
As shown in Figure~\ref{fig:tauratio}d, the axial ratios of the ellipsoids are 
generally consistent with the prolate ellipsoidal shape predicted by the model. 
Near the center of the pulse, the ratios of the ellipsoids' minor axes are 
slightly greater than one, indicating that the ellipsoids are not perfectly 
symmetric about their major axes. The slight asymmetry is most likely due to 
the nonorthogonality of the polarization modes. At phase bins 95 through 125, 
both the ellipsoid axial ratio and the intensity modulation index 
(Fig.~\ref{fig:profile}d) are relatively constant at $a_1/a_3\simeq 6.6$ and 
$\beta\simeq 0.55$, respectively, indicating that both polarization modes 
occur where the emission is heavily modulated. Throughout the pulse of PSR 
B1929+10, the modulation index is not consistent with Gaussian fluctuations in 
total intensity.

  SCRWB specifically commented on the very narrow distributions of PA and 
fractional circular polarization in PSR B1929+10. The PA distribution was
the narrowest of the pulsars they studied, and they attributed the high
degree of PA stability to an extremely homogeneous emission region. The 
supplemental information provided to SCRWB's interpretation by the eigenvalue 
analysis is any fluctuations in PA and fractional circular polarization in 
excess of the instrumental noise may be attributed to RPR in the received 
signal.

\section{DISCUSSION}
\label{sec:discuss}

  Three-dimensional measurements can assume a variety of shapes and sizes, and 
axial and eigenvalue ratios are often used to characterize these types of 
measurements (Fisher, Lewis, \& Embleton 1987). Obviously, any model that hopes 
to describe the data must be able to reproduce the fundamental shape of the 
three-dimensional data set. The shapes and axial ratios of the data point 
clusters in PSR B2020+28 and PSR B1929+10 are consistent with the predictions 
of the statistical model of radio polarimetry (McKinnon 2003b). The model 
predicts that the clusters are prolate ellipsoids when superposed OPM occurs 
and are spheroids when the polarization is fixed.  Clearly, the data are not 
consistent with some other cluster shape (e.g., dual ellipsoids or an oblate 
ellipsoid) that might be predicted by some other model (e.g., disjoint OPM). 

  In summarizing their comprehensive polarization observations, SCRWB favored 
a modal broadening mechanism as an explanation for the broad PA histograms 
they observed. However, they also allowed that the histograms could be 
explained by the superposition of OPM and RPR provided that a technique
could be developed to separate the random component from the modal emission.
Their interpretation also required the random component to be small in 
comparison to the modal emission. The analysis presented in this paper is a 
technique that allows the component separation to be made. In the analysis, 
RPR appears as the isotropic inflation of the Q-U-V data point cluster, and 
OPM manifests itself as an elongation of the cluster along a diagonal in 
Poincar\'e space. When the eigenvalue analysis is applied to the observational 
data, the random component is found to be smaller than the modal emission as 
indicated by the large axial ratios of the clusters (Fig.~\ref{fig:tauratio}). 
Expressed mathematically, the superposition of OPM and RPR represents a 
general statistical model for a polarized signal received from a pulsar.
\begin{equation}
{\rm Q} = \sin\theta_o\cos\phi_o (X_1 - X_2) + X_{\rm P,Q} + X_{\rm N,Q},
\label{eqn:Qrpr}
\end{equation}
\begin{equation}
{\rm U} = \sin\theta_o\sin\phi_o (X_1 - X_2) + X_{\rm P,U} + X_{\rm N,U},
\label{eqn:Urpr}
\end{equation}
\begin{equation}
{\rm V} = \cos\theta_o (X_1 - X_2) + X_{\rm P,V} + X_{\rm N,V}.
\label{eqn:Vrpr}
\end{equation}
Here, $\theta_o$ and $\phi_o$ are the colatitude and azimuth, respectively, of 
the OPM polarization vectors (McKinnon 2003b), and $X_{\rm P}$ is a random 
variable representing RPR. This is not to say that $X_1$, $X_2$, and $X_{\rm P}$ 
always occur at every pulse longitude of every pulsar. It is conceivable that 
any one, or even all, of these components could be absent from the received 
signal.

  Karastergiou, Johnston, \& Kramer (2003) offer a slightly different 
interpretation of the polarization fluctuations in pulsar radio emission. 
They suggest that in addition to the OPM fluctuations that occur along a 
diagonal in Poincar\'e space, the orientation of the diagonal randomly varies. 
Their interpretation does not include RPR. The Karastergiou et al. model can 
be represented by equations (\ref{eqn:Qrpr})--(\ref{eqn:Vrpr}) if the RPR 
terms, $X_P$, are ignored and if the angles $\theta_o$ and $\phi_o$ are 
allowed to be random variables.

  The phrases \lq\lq randomly polarized radiation", \lq\lq unpolarized radiation", 
and \lq\lq randomization of position angle" describe very different polarization 
processes, but they are occasionally used interchangably in the pulsar literature, 
particularly in interpretative discussions of mode-separated pulse profiles, 
single-pulse polarization observations, and radiation depolarization mechanisms. 
The differences between these processes must be explicitly described to avoid 
confusion in the use of the phrases. The Stokes parameters Q, U, and V 
recorded in an observation of an unpolarized radio source are consistent with 
instrumental noise. In Poincar\'e space, they occupy a sphere of radius 
$r=3\sigma_N$ centered on the origin (Fig.~\ref{fig:clusters}a). While RPR is 
unpolarized on average, it is not the same as unpolarized radiation. The Stokes 
parameters recorded in an observation of a randomly polarized radio source will 
also appear as a sphere centered on the origin of Poincar\'e space 
(Fig.~\ref{fig:clusters}b), but the sphere's radius will be 
$r=3(\sigma_P^2+\sigma_N^2)^{1/2}$, where $\sigma_P$ accounts for the random 
polarization fluctuations intrinsic to the source. The phrase \lq\lq randomization 
of position angle" implies fluctuations in the PA of a linear polarization vector 
which are restricted to the Q-U plane of Poincar\'e space. Polarization observations 
of a radio source with random PA fluctuations will appear as an arcing ellipsoid. 
The two largest dimensions of the ellipsoid will correspond to eigenvectors in the 
Q-U plane. The ellipsoid's centroid will be offset from the origin by a distance 
equal to the polarization of the source.

  The Gaussian intensity fluctuations and the random polarization observed 
at the center of the PSR B2020+28 pulse suggest a possible origin of RPR. 
Pulsar radio emission has been described as an incoherent addition of 
individually coherent shot pulses (Rickett 1975; Cordes 1976a, 1976b; Hankins 
et al. 2003). The addition can occur within the telescope receiving system, 
in the dispersion and scattering of the pulses as they propagate through the 
ISM, and at the pulsar if many independent emission events contribute to the 
instantaneous signal (Cordes 1976a, 1976b). From the Central Limit Theorem, 
the recorded signal will have Gaussian statistics regardless of the intensity 
distribution of the shot pulses, provided the number of pulses in the incoherent 
sum is very large. If the shot pulses also have no preferred polarization, the 
polarization of the recorded signal will be completely random. This scenario 
suggests that the intensity and polarization of the RPR component ($X_{\rm P}$ 
in eqns. (\ref{eqn:Qrpr})--(\ref{eqn:Vrpr})) always follow Gaussian statistics. 
It is difficult to extract the statistics of $X_{\rm P}$ in PSR B1929+10 and 
at other locations within PSR B2020+28 because of the large polarization 
fluctuations due to OPM. Cairns et al. (2003) have found evidence for a small 
Gaussian emission component in the Vela pulsar using theoretical models of its 
longitude-dependent intensity distributions. They also attribute the Gaussian 
component to the incoherent sum of shot pulses. 

  The eigenvalue analysis may be used to determine how pulsar radio emission 
depolarizes at high frequency (Manchester, Taylor, \& Huguenin 1973; Morris, 
Graham, \& Sieber 1981). The depolarization has been attributed to OPM, 
randomization of PA, a process intrinsic to the emission mechanism, or any 
combination of the three (MTH; SCRWB; Xilouris et al. 1994; McKinnon 1997; MS1;
Karastergiou et al. 2002). Since the observed Q-U-V clusters are not the arcing 
ellipsoids expected for fluctuations in PA, the analysis presented here suggests 
that randomization of PA does not occur. Therefore, randomization of PA is not 
a strong candidate for a depolarization mechanism. However, the analysis 
indicates that RPR occurs in the received pulsar signal, and RPR may be a 
viable depolarization mechanism if it is generated at emission. Applying the 
eigenvalue analysis to high frequency polarization observations of pulsars may 
determine if OPM, RPR, or both lead to the depolarization. OPM will be the 
depolarization mechanism if the high frequency Q-U-V clusters are highly 
elongated ellipsoids with dimensions of their minor axes set by the 
instrumental noise. Since the radiation is depolarized, the cluster centroid 
should be located near the origin of Poincar\'e space. If RPR depolarizes the 
emission, the high frequency Q-U-V clusters will be spheroids with radii much 
larger than the instrumental noise and with centroids near the Poincar\'e 
origin. Applying the eigenvalue analysis to multi-frequency, polarization 
observations of pulsars can also be used to measure the spectrum of RPR and 
the longitudinal extent of OPM, both of which will be useful in determining 
the depolarization mechanism. 

  The presence of RPR in the received signal has other subtle, yet important, 
consequences for pulsar polarization observations and our interpretation of 
them. Perhaps the most important consequence of RPR is that it masquerades as 
on-pulse instrumental noise, thereby causing histograms of linear polarization,
circular polarization, polarization position angle, fractional linear 
polarization, and fractional circular polarization to be much wider than what 
is expected from the off-pulse instrumental noise. McKinnon (2002) derived the 
statistics of fractional polarization in an attempt to explain the large 
dispersion in fractional circular polarization reported by SCRWB. The excess 
dispersion was attributed to a combination of the signal-to-noise ratio of 
the observations, the heavy modulation of the mode intensities, and the small 
degree of circular polarization intrinsic to OPM. However, the derivation did 
not account for RPR, which increases the effective noise and thus decreases 
the effective signal-to-noise ratio. The factor of two to three increase in 
effective noise indicated by the eigenvalue analysis of PSR B2020+28 and PSR 
B1929+10 is enough to attribute the observed dispersion in fractional circular 
polarization to the reduced signal-to-noise ratio caused by RPR. Similarly,
Kramer et al. (2002) found that the distribution of fractional circular
polarization in the Vela pulsar was much wider than what was expected from 
the off-pulse instrumental noise. RPR could account for the excess noise 
in the emission from Vela and, as mentioned above, may be the Gaussian 
emission component described by Cairns et al. (2003).  McKinnon \& 
Stinebring (2000) attributed excess polarization fluctuations in PSR B2020+28 
to on-pulse instrumental noise, but their noise model could not account for 
all the observed fluctuations. The RPR revealed by the eigenvalue analysis can 
easily account for the excess fluctuations. RPR also causes the linear 
polarization to be slightly overestimated. When reporting the results of 
pulsar polarization observations, it is common practice to subtract a bias 
offset caused by the off-pulse instrumental noise from the computed values 
of linear polarization. While the bias value of 
\begin{equation}
\langle L\rangle =\sigma_N\sqrt{\pi/2}
\end{equation}
is relevant for data recorded off the pulse, the actual value may be much 
larger on the pulse where RPR can occur. 
\begin{equation}
\langle L\rangle = \sqrt{(\sigma_P^2+\sigma_N^2)\pi/2}
\end{equation}

  The OPM statistical model requires fluctuations in mode polarization
amplitudes to observe the switching between modes. Similarly, moding will 
not be observed if the polarization amplitudes do not fluctuate. If the 
polarization amplitude fluctuates, one might reasonably expect the total 
intensity to fluctuate, also (MS1; MS2). Consequently, the model predicts 
that OPM should occur where the radiation is heavily modulated 
(\S~\ref{sec:opm}). This is clearly the case for PSR B2020+28 and PSR 
B1929+10, as shown in Figures~\ref{fig:profile} and~\ref{fig:tauratio}. 
The presence of OPM elongates the polarization ellipsoid, giving it a large 
axial ratio. For both pulsars, the axial ratio is large where the modulation 
index is also high. The axial ratio in PSR B1929+10 is relatively constant 
across much of its pulse, and its modulation index remains constant over a 
similar range in pulse longitude. In PSR B2020+28, both the axial ratio of 
the polarization ellipsoid and the intensity modulation index are large at 
the same locations within the pulse, primarily in the pulse peaks and wings. 
As mentioned in \S~\ref{sec:2020}, the intensity fluctuations are Gaussian 
and the modulation index is $\beta < 0.3$ near the center of the PSR B2020+28 
pulse. The polarization data in this region are consistent with the occurrence 
of only one polarization mode. This observation raises the intriguing 
possibility that both polarization modes do not occur where intensity fluctuations 
are Gaussian. Stated another way, perhaps both polarization modes occur only 
where the modulation index exceeds some critical value, $\beta_c\simeq 0.3$.
  
  The longitudinal extent of OPM determined from the eigenvalue analysis 
(Fig.~\ref{fig:tauratio}) is different from what has been determined in 
previous analyses. MS2 calculated the mode-separated pulse profiles of
PSR B0525+21 and PSR B2020+28. The model used in their analysis neglected 
RPR, and their results indicate that both modes occur at all locations 
within the pulse of each pulsar. However, the eigenvalue analysis presented 
in this paper reveals the presence of RPR and suggests that only one mode 
occurs near the center of the PSR B2020+28 pulse. Using PA histograms,
SCRWB found that OPM extended over a limited range of pulse longitudes. 
The elongated Q-U-V clusters may be a better indicator of OPM because, 
like the bin 90 data in PSR B1929+10, the PA histograms may only reveal 
the presence of one mode although both modes actually occur. Therefore, 
the longitudinal extent of OPM may be slightly greater than what is 
documented in SCRWB.

  RPR complicates the mode separation technique presented in MS2. The
technique assumes that two superposed modes, each completely polarized,
are the only components of the emission. The simplicity of this assumption
allows the mode intensities and polarizations to be calculated from the 
average values of the measured Stokes parameters. The occurrence of RPR
introduces another variable in the model (e.g. Eqns.~\ref{eqn:Qrpr} 
-~\ref{eqn:Vrpr}), and the mode separation is then indeterminate unless 
some simplifying assumption regarding RPR can be made. The dominant effect 
of RPR upon the results presented in Figures 4 and 5 of MS2 will be to 
reduce the intensities and polarizations of the modes at all pulse 
longitudes. This effect may be small because the magnitude of RPR 
estimated in this analysis appears to be small in relation to the 
total intensity of the emission (Fig.~\ref{fig:tauratio}).

  The results of single-pulse polarization observations are notoriously difficult 
to summarize and report, owing to the extreme variability of pulsar polarization 
and the large volume of data. The analysis developed here is a new method for 
presenting and interpreting the results of these types of observations. It is 
significantly different from previous data presentation methods. The results of
some of the first single-pulse polarization observations were reported by plotting
the raw Q-U-V data for a series of single pulses (e.g., Lyne, Smith, \& Graham 
1971). These seminal observations showed that pulsar polarization can be quite 
high and highly variable. Although the orientation of the polarization vector is 
somewhat difficult to visualize with this data presentation method, it readily 
illustrates the polarization associated with the total intensity of a limited 
number of pulses. MTH reported the results of their single-pulse polarization 
observations as multiple polarization ellipses plotted across the total intensity 
profiles of about a dozen, consecutive, individual pulses. The radiation's 
polarization state is immediately obvious with the MTH data presentation method.
These observations were some of the first to reveal conclusively that OPM
occur in the emission. SCRWB and Backer \& Rankin (1980) presented the results 
of their single-pulse polarization observations as histograms of fractional 
circular polarization, fractional linear polarization, and PA. The histograms, 
which are particularly well-suited for documenting the occurrence of OPM, are 
statistical summaries that conveniently illustrate how the polarization 
fluctuates across the pulse. The polarization data recorded for a large number 
of pulses can be reported with this method. However, the histograms do not 
indicate the instantaneous orientation of the radiation's polarization vector 
because the circular and linear polarization data are reported separately. The 
data presentation method developed in this paper focuses on a detailed 
investigation of a very large number of polarization samples at a single pulse 
longitude at the expense of both the total intensity information and a more 
global, instantaneous view of the polarization at other pulse longitudes. The 
method immediately shows the orientation of the radiation's polarization vector, 
reveals the presence of OPM through the elongation of the data point cluster, 
and indicates the presence of RPR through the inflation of the cluster. Data 
presented in this manner can be readily interpreted in the context of other 
mechanisms that cause polarization fluctuations. 

  The objective of most radio polarization observations is to determine where
the tip of a polarization vector resides in Poincar\'e space. This is done
simply by averaging the individual Stokes parameters Q, U, and V. The tip of 
the vector is the centroid of the Q-U-V data point cluster, and it can be 
located more precisely by reducing the cluster size (i.e. the instrumental 
noise) through increased bandwidth and longer integration times. But as shown 
with the eigenvalue analysis and the statistical model of radio polarimetry
(McKinnon 2003b), more polarization information is available to the radio 
astronomer. The unique ability to measure all Stokes parameters simultaneously 
at radio wavelengths also allows the astronomer to determine the cluster's 
size, shape, and orientation. The physical processes that create polarization 
fluctuations can be investigated by measuring all geometrical properties of 
the cluster.

\section{CONCLUSIONS}
\label{sec:conclude}

  In summary, an eigenvalue analysis has been developed for the interpretation
of pulsar polarization data. The analysis is based upon a three-dimensional 
statistical model of radio polarimetry and allows the linear and circular
polarization of pulsar radio emission to be interpreted together, instead of 
separately as has been done typically in the past. The model is generally 
consistent with polarization observations of PSR B1929+10 and PSR B2020+28. 
The analysis clarifies the origin of polarization fluctuations in pulsar radio 
emission. The observed excess dispersion in PA is caused by the isotropic 
inflation of the Q-U-V data point cluster. The cluster inflation is not 
consistent with random fluctuations in PA and is attributed to RPR in the 
received pulsar signal. The presence of RPR may contribute to the 
depolarization of the emission at high radio frequency. A comparison of the 
model with the observational data suggests that OPM are superposed. OPM occur 
where the emission is heavily modulated, and may only occur where the 
modulation index exceeds a critical value, $\beta_c\simeq 0.3$. The analysis 
also reveals deviations from mode orthogonality in the circular polarization 
of PSR B1929+10. 

\acknowledgements
I thank Dan Stinebring for providing the data used in the analysis. I thank 
Tim Hankins and an anonymous referee for constructive comments on the paper. 
I am grateful for the hospitality of the NRAO staff in Green Bank, WV, where 
this paper was written. The Arecibo Observatory is operated by Cornell 
University under cooperative agreement with the National Science Foundation.


\end{document}